# Nucleobase-functionalized graphene nanoribbons for accurate high-speed DNA sequencing

Eugene Paulechka[1], Tsjerk A. Wassenaar[2,3], Kenneth Kroenlein[1], Andrei Kazakov[1], and Alex Smolyanitsky[1*]

[1]Applied Chemicals and Materials Division, National Institute of Standards and Technology,

Boulder, CO 80301, USA

[2]Department of Biology, Friedrich-Alexander University of Erlangen-Nürnberg,

91058 Erlangen, Germany

[3]Groningen Biomolecular Sciences and Biotechnology Institute and Zernike Institute for Advanced Materials, University of Groningen,

9747 AG Groningen, the Netherlands

[*]To whom correspondence should be addressed: alex.smolyanitsky@nist.gov

## Abstract

**We propose a water-immersed nucleobase-functionalized suspended graphene nanoribbon as an intrinsically selective device for nucleotide detection. The proposed sensing method combines Watson-Crick selective base pairing with graphene's capacity for converting anisotropic lattice strain to changes in an electrical current at the nanoscale. Using detailed atomistic molecular dynamics (MD) simulations, we study sensor operation *at ambient conditions*. We combine simulated data with theoretical arguments to estimate the levels of measurable electrical signal variation in response to strains and determine that the proposed sensing mechanism shows significant promise for realistic DNA sensing devices without the need for advanced data processing, or highly restrictive operational conditions.**

Fast, reliable, and cost-effective DNA sequencing continues to be an important open problem, as the vast majority of sequencing needs currently remains to be satisfied by the use of the Sanger method [1]. Despite the numerous advances made in automation and data analysis as part of the Human Genome Project [2, 3], the throughput rate and cost still significantly limit routine production of genomic data using the currently available technology.

Various alternative approaches have been proposed, ranging from employing ionic current blockage by DNA nucleotides in aqueous nanopores [4, 5] to the use of chemically selective tunneling current probes [6]. Discovery of atomically thin carbon allotropes and their exceptional electronic, mechanical, and chemical properties has reinvigorated the search for nanoscale system based DNA sequencing methods in the past decade. Consequently, numerous graphene-based approaches have been proposed, mostly centered on the use of graphene as the ultimately thin membrane impermeable to water-dissolved ions [7-10] and nanoscale graphene-based field-effect transistor devices with nucleotide-specific electronic response [11-15]. In all proposed methods, robust single-measurement nucleobase selectivity in realistic measurement conditions naturally remains a fundamental challenge [16, 17]. with single nucleotide count errors of up to 90% [18], depending on the approach. In addition, device noise in ambient conditions remains one of the most serious problems in developing a robust graphene-based sensing methodology [12].

Here, we report on utilizing graphene's electronic properties, effectively combined with the Watson-Crick base-pairing, as a possible method of high-speed DNA sequencing at ambient conditions in aqueous environment. The key feature of the proposed method is a graphene nanoribbon (GNR) with a nanoscale opening, the interior of which is chemically functionalized with selected nucleobases. As sketched in Fig. 1 (a), a single-strand DNA (ssDNA) molecule is inserted into the functionalized pore and translocated at a prescribed rate perpendicularly to the GNR. When a base complementary to the GNR's functional group traverses the pore during translocation, selective hydrogen bond formation is expected to occur, as shown in Fig. 1 (b), provided there is sufficient dwell time set by the prescribed translocation rate. As a result of local DNA-GNR binding, the GNR is expected to be temporarily pulled upon and deflected in the out-of-plane direction, followed by a slip when the critical force required for breaking the hydrogen bonds is reached. Although direct microscopy-based methods of detecting such deflections are possible, here we focus on the possibility of utilizing the effect of deflection-induced strain on the electronic properties of the GNR, and thus detecting a temporary change in the electrical current (from the electrical bias, as sketched in Fig. 1 (a)), as discussed later. Therefore, temporary changes in the electrical current can serve as electrically measurable nucleobase detection events. Given that C/G-functionalized GNR is a G/C-selective, while an A/T-functionalized GNR is T/A-selective, vertically stacking a total of four independently biased and appropriately functionalized GNRs would result in an integrated sequence detector.

Here, we present the results of carefully designed atomistic molecular dynamics (MD) simulations of the continuous ssDNA translocation through a C-functionalized GNR in an aqueous environment. Our simulations are aimed at assessing the selectivity of detecting G nucleobases in terms of the effective GNR deflection at room temperature.

Shown in Fig. 2 (a) is the central portion of the sensor, which consists of a $L_x = 4.5$ nm $\times L_y = 15.5$ nm GNR with a ~2.5 nm -wide nanopore, whose interior is functionalized with three cytosine molecules. The GNR is position-restrained at the ends, mimicking suspension between solid electrodes. The carbon atom at position six in the cytosine molecule was covalently attached to an edge carbon [19] in the pore of the functionalized GNR (FGNR), as seen in the magnification in Fig. 2 (a). Such an attachment point keeps the hydrogen-bonding groups facing the interior of the nanopore, and thus available for interaction with the side-chains of the nucleotides subject to sensing. The atomic-level geometry of the cytosine-functionalized region, including the near-90° orientation (consistent with previous calculations [20]) of the cytosine moiety relative to the graphene plane, was obtained from a DFT energy optimization of an anthracene molecule, functionalized by the cytosine moiety at position nine. As shown in Fig. 2 (b), six-residue ssDNA samples (sequences defined further) periodic in the Z-direction was inserted into the nanopore, as shown in Fig. 2 (c). All DNA backbones were pre-stretched in the Z-direction by a force of ~0.1 nN in order to promote strand linearity. In addition, weak positional restraint in the XY-plane was applied to the ssDNA backbone, so it on average remained close to the center of the nanopore, essentially mimicking a precision aperture. All MD simulations of the DNA translocation process were performed at ambient conditions ($T = 300$ K, $p = 0.1$ MPa). The DFT optimization of the GNR functional region was performed with the use of Gaussian 09[21] at the *B3LYP/6-31+G(d)* theory level [22, 23]. All partial atomic charges in the functional group were calculated according to the CHELPG scheme [24] at the *HF/6-31+G(d)* theory level for the optimized geometry. The MD models of the ssDNA and the FGNR were based on the OPLS-AA forcefield [25, 26]. The intramolecular interaction parameters for graphene were obtained from the optimized bond-order potential for carbon [27], as described elsewhere [28]. As shown in Fig. 2 (c), the entire system was immersed in a rectangular box filled with water described by the TIP4P model [29, 30]. Prior to the production MD runs, the systems were carefully pre-relaxed in *NPT* simulations at $T = 300$ K and $p = 0.1$ MPa, while the production simulations of DNA translocation were performed in an *NVT* ensemble at $T = 300$ K, using GROMACS v. 5.0.5 software [31, 32].

The results obtained in a simulated ssDNA translocation aimed at sensing the G residues by a C-functionalized GNR are discussed next. We used two arbitrarily selected six-residue periodic sequences of *GAAGCT* (SEQ1) and *TCGAAC* (SEQ2) translocated in the negative-Z direction at a constant rate of 5 cm/s, as dictated by the MD time limitations for a system of this size. The FGNR in all cases was pre-stretched at the ends (along the Y-axis) by 0.5 % to enable a rapid return to post-deflection unperturbed state and to somewhat suppress the thermal fluctuations. The total simulated time was 300 ns, during which the sequences translocated through the pore were repeated approximately 3.3 times. In the case of SEQ1, this effectively corresponds to *GAAGCT*/*GAAGCT*/*GAAGCT*/*GA*, and thus seven passes of the G residue are expected. For SEQ2, the corresponding sequence is *TCGAAC*/*TCGAAC*/*TCGAAC*/*TC* with three expected passes of G. Note that the nucleobase inside the pore at the start of a simulation (underlined) can vary, because the pre-translocation MD relaxation steps for the various systems discussed here allowed for spurious translocation of the sample ssDNA. As seen further, the variation in the starting residue does not affect the clarity of our observations, as we effectively track the passage of the G-bases. Shown in Fig. 3 (a) we present the results of SEQ1 translocation in the form of the out-of-plane deflection of the GNR as a function of simulated time, shown alongside the number of hydrogen bonds formed between the functional groups and the G residues in the tested ssDNA. The latter measure enables clear tracking of the residue of interest, as it passes through the pore. As expected, a total of seven binding events occur between the functional groups at the pore interior and the residues of interest. A total of three hydrogen bonds are created in each case, as shown in Fig. 3 (a), consistent with the schematic representation in Fig. 1 (b). Remarkably, in *six out of the seven* G-passage events, the bond formation is accompanied by a clear GNR out-of-plane deflection of the average ~2 Å magnitude, followed in each case by a sharp slip when the critical force required for breaking the G-C hydrogen bonds was reached. The clearly missed event is located at $t \approx 100$ ns. Interestingly, as observed in Fig. 3 (e), the critical pulling force in the range 50-80 pN, consistent with the experimentally measured 54.0-61.4 pN [33, 34]. Note that for a sensor aimed at detecting the A/T bases (with a T/A-functionalized GNR), the corresponding critical force would be ~2/3 of the average C-G value, reducing the corresponding deflection magnitude accordingly (see Supplementary Information). In Figs. 3 (b, d, f), the same data is presented for SEQ2. Similarly, the passage of the G-nucleobase is accompanied by a temporary GNR deflection *in all of the three cases*. It is noteworthy that no post-processing of the deflection data was performed beyond basic low-pass filtering with a bandwidth of 500 MHz, as stated in the figure legend. Such filtering allowed effective removal of thermal noise, as well as clear resolution of the deflection events, which, from the presented data, took ~15 ns on average, thus corresponding to an effective frequency of ~66 MHz.

The intended ssDNA-FGNR chemical coupling can be seen in Fig. 4, where a representative deflected state of the FGNR immediately prior to pull-off is shown. The interacting groups are properly oriented and the hydrogen bonds are formed between the functional group at the nanopore edge and the passing G-base. Importantly, no false-positives occurred during simulated translocation, likely due to no significant hydrogen bond formation between the FGNR and the non-guanine nucleobases in the test sample. This cause for the demonstrated level of selectivity is supported by the data in Figs 3 (c, d), where all "spurious" hydrogen bonds are tracked throughout the simulated time alongside the intended (FGNR-G) bonds.

The translocation rate is expected to greatly affect the device performance in terms of the bond formation and the overall system relaxation, resulting in effects on the noise, and nucleobase selectivity. Therefore, we performed translocation tests of SEQ1 at a significantly higher rate of 25 cm/s, corresponding approximately to 330 million nucleobases per second, simulated for 60 ns. The resulting FGNR deflection and the number of C-G hydrogen bonds as functions of time are shown in Fig. 5. Although C-G bonds continue to form between the FGNR and the passing ssDNA, and the average deflection magnitude is higher (likely due to lack of system relaxation), the detection quality here is noticeably worse. In particular, we observe a noisy resolution of the G passage at $t \approx$ 10 ns and a missed event at $t \approx$ 35 ns.

Although one cannot claim true statistical significance from the results of atomistic simulations describing translocation of only ten sequencing events, our combined data in Figs. 3 (a, b) suggest an overall *single-sensor* detection error probability in the vicinity of 1/10 for the 5 cm/s translocation rate. Given that no false positives were observed, one roughly estimates that a total of *four independent measurements* would be required to achieve a 99.99 % fidelity at the translocation rate of 5 cm/s. In addition, given that an experimental setup would likely use even lower translocation speeds, the overall *single-measurement* error rate may be further improved.

As mentioned earlier, the nucleotide passage can be in principle detected in an atomic force microscopy-like setup by tracking the deflections directly, or by monitoring the FGNR-DNA interaction forces (shown in Figs. 3 (e, f)). However, direct electronic detection of the nanoscale deflection is a highly attractive option. Although a truly accurate estimate of the electrical current changes in a transversely deflected FGNR would depend on the GNR quality, edge geometry (zigzag or armchair), and dimensions, a theoretical discussion at the order-of-magnitude level is possible. Electronic deflectometry based on the effect of strain on the electronic band structure of carbon nanotubes and graphene has been studied in detail experimentally[35] and theoretically [36, 37]. Assuming negligible contact resistance and positing that thermally activated carriers in the conduction band dominate at $T = 300$ K, the Landauer formalism yields a relative change in electrical resistance of a GNR of the order $\frac{\Delta R}{R} \approx \frac{\Delta E_{gap}}{kT}$, [35] where $R$ is the electrical resistance, $\Delta E_{gap}$ is the energy bandgap modulation at the Dirac point due to the deflection-induced uniaxial strain (regardless of the existing bandgap in an undeflected state due to GNR edge type, width, *etc*) and *k* is the Boltzmann constant. A tight-binding estimate for the effect of uniaxial strain $\varepsilon$ is $\Delta E_{gap} \propto 3t_0\varepsilon$ ($t_0 \approx 2.7\ eV$ is the nearest-neighbor electron hopping energy for graphene).[38] For a GNR of length $L$ deflected by $h \ll L$, $\varepsilon \approx 2(h/L)^2$, and thus for the FGNR dimensions and $h$ = 2 Å in this work, we estimate $\varepsilon = 0.033$ %, yielding a positive $\boldsymbol{\frac{\Delta R}{R} = 10.4\%}$ due to the largeness of the $(t_0/kT)$ ratio, and thus resulting in an appreciable *average current decrease*.

If, depending on the GNR, coherent transport dominates the conductive process, the effect of strain on an ungated GNR would be negligible [36]. Thus, an effective use of a gate electrode was suggested [36], which enables modification of the carrier transmission probability and results in an effective relative change in conductance of the order $(h^2/La_0)$, where $a_0 = 1.42$ Å is the C-C bond length in graphene. The latter estimate yields a relative change of $\approx 2$ % for the deflected GNR considered here. This net change is not particularly high, but is experimentally detectable and is higher in larger GNRs. Recall that this work considers a short and narrow GNR (dictated by the computational limitations of MD). In an experimental setup, a considerably wider and longer GNR would be used. The deflection-to-length ratio is $\frac{h}{L} \sim (F_c/w)^{1/3}$, where *w* is the GNR width and $F_c$ is the critical C-G shearing force (see Supplementary

Information), and thus the amount of deflection could be increased in a longer and wider GNR with a somewhat higher aspect ratio. For instance, a 10 nm × 60 nm GNR would be deflected (at $F_c = const$) by ≈ 6.4 Å, thus resulting in a relative average current shift by ≈ 4.8%. In a GNR with the original dimensions and *without lateral pre-strain*, the two detection mechanisms discussed above yield a relative change in resistance of 71 % and 12 %, respectively. See sections S1 and S2 of Supplementary Information for details, as well as further discussion on the GNR size effects.

Because the locally suspended FGNR is part of an aqueous system, subject to interaction with water and the nanopore-confined DNA molecule, the effect of dynamic ripples on the GNR can introduce an additional source of charge carrier scattering and noise, compared to "conventional" solid-state devices. As estimated in the Supplementary Information, the ripple scattering strength is decreased as a result of deflection-induced strain. Therefore, if ripple scattering is expected to significantly contribute to the overall resistance in a given GNR, the observed effect may become a contributing mechanism, suggesting an additional design consideration in terms of the GNR dimensions, edge type, and doping parameters. The rippling process is intrinsically dynamic in suspended atomically thin membranes, causing significant rippling mean-square variation in the time domain (see Fig. S1 of the Supplementary information). This suggests a temporal modulation of the local electron hopping parameters, and thus an additional source of rippling-induced noise. However, because the timescale of the ripple dynamics is fundamentally linked to the flexural wave propagation velocity in graphene of the order of kilometers per second [39], the dynamic modulation of current occurs at the picosecond timescale [40], unless the effective GNR dimensions reach tens of microns. Thus, given that the timescale of the deflection-induced signal is of the order of tens of nanoseconds for the translocation rate of 5 cm/s (~15 ns/base), and expected to be further lowered in an experiment (currently 1-3 µs/base [41]), one can argue that basic low-pass signal filtering should be sufficient to eliminate the high-frequency current noise arising from the FGNR fluctuations.

As a result of its hydrophobicity, graphene is likely to adsorb a DNA strand in the case of lack of positional precision during insertion and translocation in an experiment, suggesting a significant effect on the overall measurement accuracy. A possible way to alleviate this potential problem is by locally tailoring graphene hydrophobicity in the vicinity of the nanopore via non-covalent coating [42], expected to also affect the GNR flexibility and thus the amount of strain-inducing deflection. In addition, the concept outlined here is extendable toward any non-hydrophobic atomically thin membranes with sufficient sensitivity of the electronic properties to anisotropic strain, *e.g.* molybdenum disulfide [43].

**Conclusions**

We have proposed a cytosine-functionalized graphene nanoribbon deflectometer *in aqueous environment at ambient conditions* for fast and accurate sensing of nucleotides during continuous translocation of intact ssDNA. The proposed sensing mechanism combines Watson-Crick complementary base pairing with the sensitivity of graphene's electronic properties to anisotropic strain, as well as the deflection-induced field effects described in the literature. Our simulations demonstrate single-sensor guanine detection accuracy in the vicinity of 90 % with no false-positives at the translocation rate of *~66 million nucleobases per second*. We estimate that the strain effects on the electrical conductance of graphene nanoribbons at ambient conditions are measurable, requiring only basic low-pass signal filtering, and thus rapid sequencing may be effectively reduced to an electrical current measurement in the milliampere range (as dictated by a typical electrical resistance of a GNR [13]) without the need for microscopy-based methods. With the exception of the proposed nucleobase functionalization, all of the individual components of the proposed sensing method have been previously implemented experimentally. The proposed sensing methodology may therefore hold significant promise for realistic DNA sequencing devices without the need for advanced data processing, or highly restrictive operational conditions, especially given its extendibility toward other strain-sensitive membranes.

**Authors' contributions**

A.S. conceived the sensor concept and supervised the project. E.P. developed and detailed GNR functionalization, and devised DFT simulations. A.S. and E.P. designed MD simulations and analyzed the data. T.A.W. built the DNA model and automated system design. T.A.W. and A.S. performed MD simulations. K.K. devised and implemented data analysis software. A.K. performed DFT simulations and contributed to the data analysis software. All authors discussed the results, composed the manuscript, and contributed to revisions.

**Acknowledgment**

The authors are grateful to A. Isacsson and D.K. Ferry for illuminating discussions. T.A.W. acknowledges support by S.J. Marrink and the generous computational support by the Donald Smit Center at the University of Groningen. A.S. acknowledges the computational support by the NIST Center for Theoretical and Computational Materials Science and personally thanks A.C.E. Reid. We also thank L. Rast for providing outstanding assistance with graphics.

**Figures**

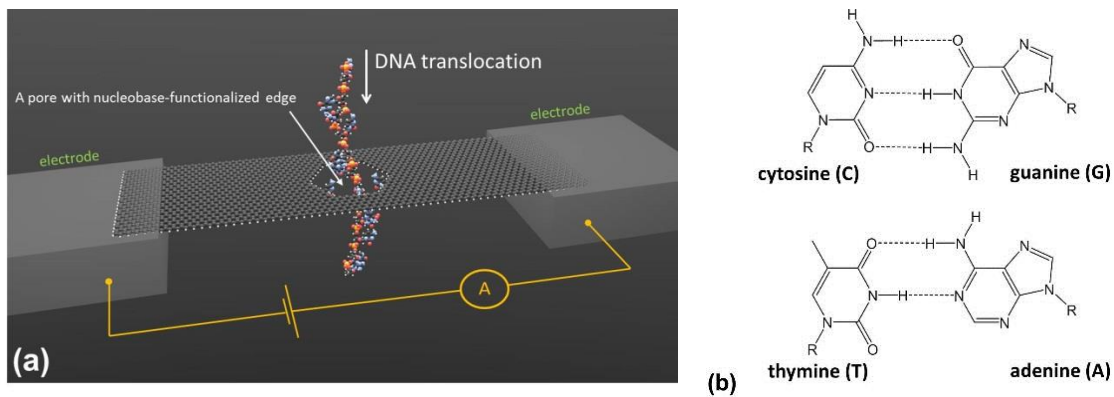

Figure 1. A three-dimensional sketch of the proposed nucleobase sensor (a) and the Watson-Crick base-pairing principle (b). The dotted lines in (b) represent the hydrogen bonds formed between the nucleobases.

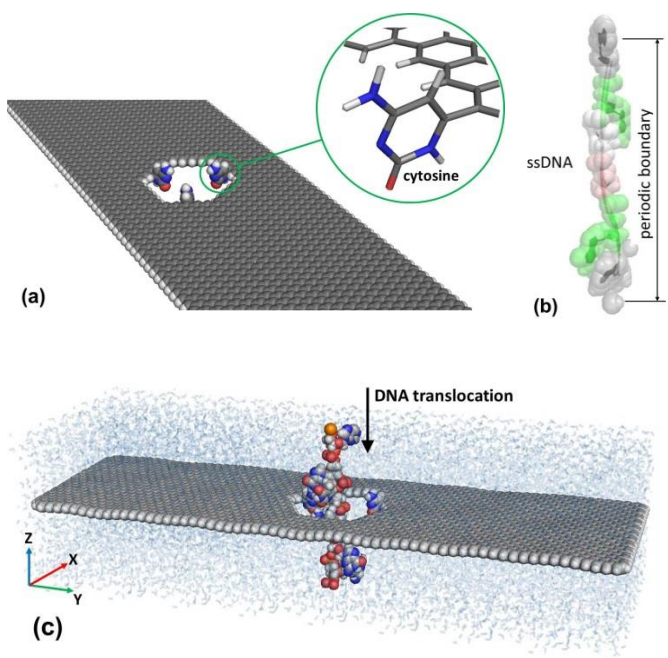

Figure 2. An atomistic model of a graphene nanoribbon with cytosine-functionalized nanopore (a), periodic ssDNA (b), and the complete simulated assembly immersed in water (c).

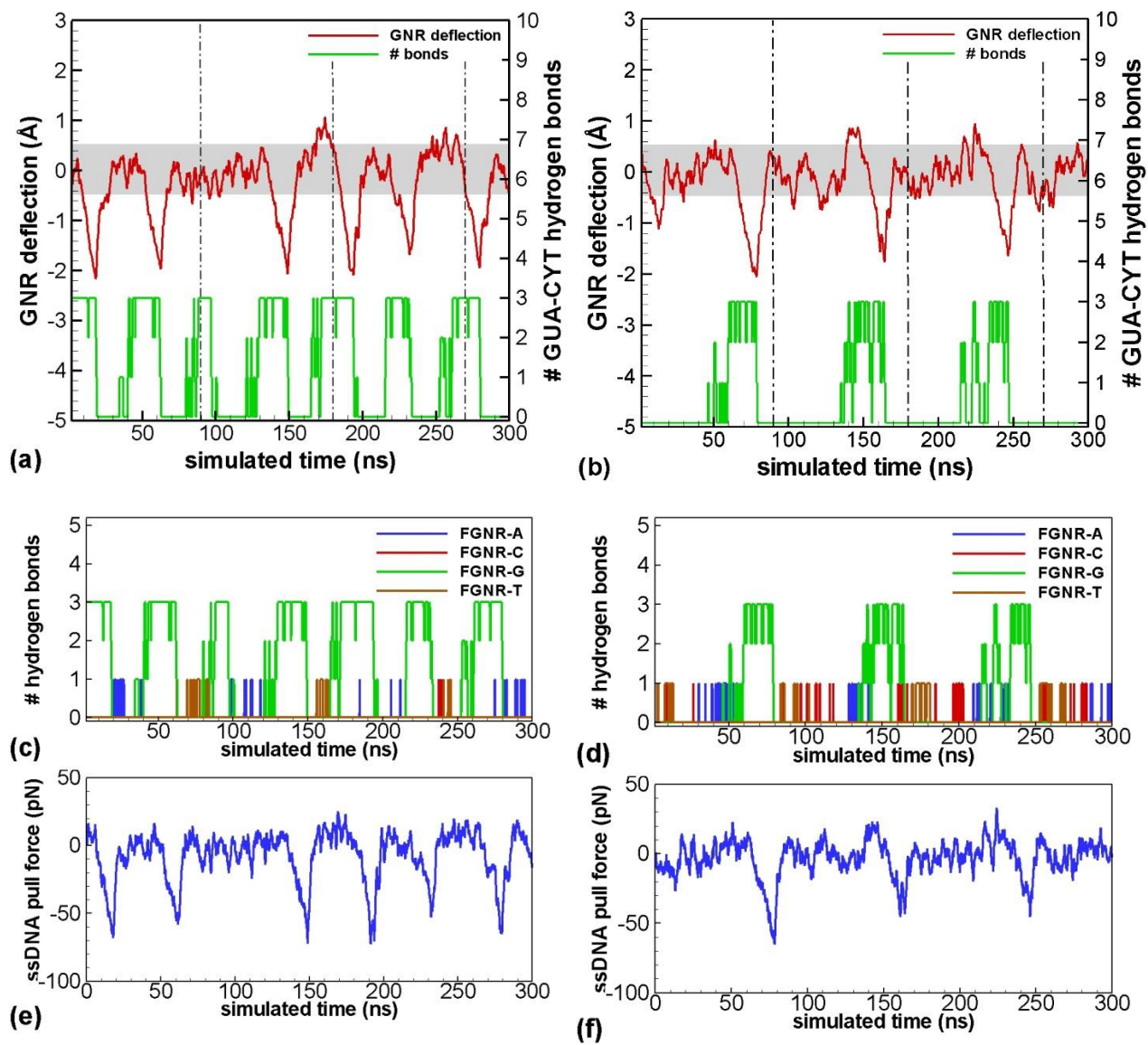

Figure 3. FGNR deflection and the number of G-C hydrogen bonds as functions of simulated time for ssDNA sequence SEQ1 (periodic *GAAGCT*) (a) and sequence SEQ2 (periodic *TCGAAC*) (b); all hydrogen bonds between the FGNR and the ssDNA sequence SEQ1 (c) and SEQ2 (d); ssDNA pulling force during translocation of SEQ1 (e) and SEQ2 (f). The ssDNA translocation speed in these simulations was 5 cm/s, approximately corresponding to 66 million of nucleobases per second. The grey strips in (a) and (b) represent a background noise measure: the strip half-width is $\sqrt{2} \times$ deflection RMSD from the regions of non-G translocation in (b). A low-pass filter with an effective bandwidth of 500 MHz was applied to the raw deflection and pulling force data.

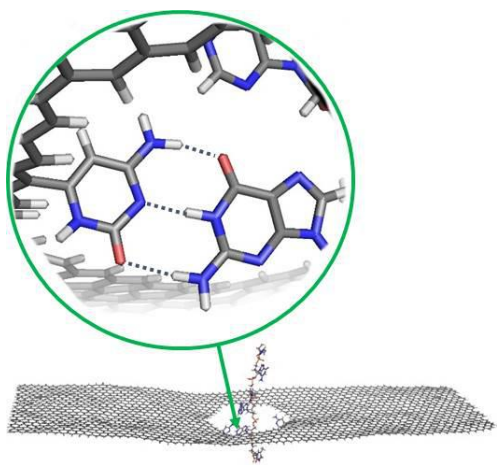

Figure 4. A representative deflected state of the FGNR prior to G-C detachment.

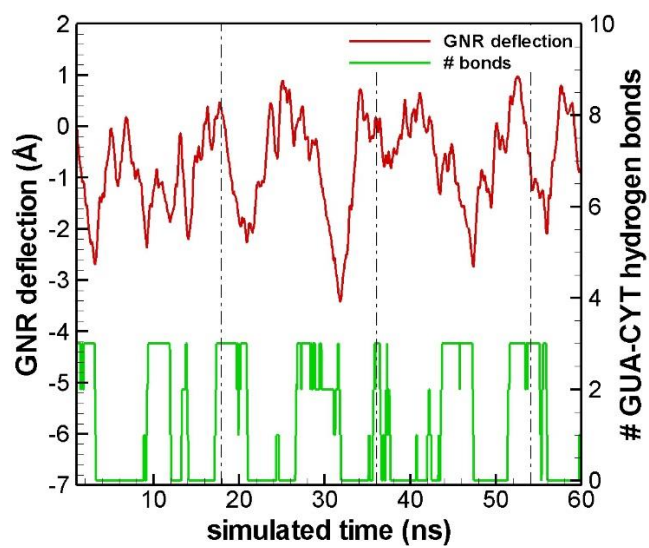

Figure 5. FGNR deflection and the number of G-C hydrogen bonds as functions of simulated time during ssDNA (SEQ1) translocation at the rate of 25 cm/s. A low-pass filter with an effective bandwidth of 2.5 GHz was applied to the raw deflection data.

# References

1. Sanger, F., S. Nicklen, and A.R. Coulson, *DNA sequencing with chain-terminating inhibitors.* PNAS, 1977. **74**(12): p. 5463-5467.
2. Shendure, J., et al., *Advanced sequencing technologies: methods and goals.* Nat Rev Genet, 2004. **5**(5): p. 335-344.
3. Shendure, J. and H. Ji, *Next-generation DNA sequencing.* Nat Biotech, 2008. **26**(10): p. 1135-1145.
4. Kasianowicz, J.J., et al., *Characterization of individual polynucleotide molecules using a membrane channel.* Proceedings of the National Academy of Sciences, 1996. **93**(24): p. 13770-13773.
5. Harrell, C.C., et al., *Resistive-Pulse DNA Detection with a Conical Nanopore Sensor.* Langmuir, 2006. **22**(25): p. 10837-10843.
6. He, J., et al., *Identification of DNA Basepairing via Tunnel-Current Decay.* Nano Letters, 2007. **7**(12): p. 3854-3858.
7. Garaj, S., et al., *Graphene as a subnanometre trans-electrode membrane.* Nature, 2010. **467**(7312): p. 190-193.
8. Wells, D.B., et al., *Assessing Graphene Nanopores for Sequencing DNA.* Nano Letters, 2012. **12**(8): p. 4117-4123.
9. Schneider, G.F., et al., *DNA Translocation through Graphene Nanopores.* Nano Letters, 2010. **10**(8): p. 3163-3167.
10. Merchant, C.A., et al., *DNA Translocation through Graphene Nanopores.* Nano Letters, 2010. **10**(8): p. 2915-2921.
11. Min, S.K., et al., *Fast DNA sequencing with a graphene-based nanochannel device.* Nat Nano, 2011. **6**(3): p. 162-165.
12. Dontschuk, N., et al., *A graphene field-effect transistor as a molecule-specific probe of DNA nucleobases.* Nat Commun, 2015. **6**.
13. Nelson, T., B. Zhang, and O.V. Prezhdo, *Detection of Nucleic Acids with Graphene Nanopores: Ab Initio Characterization of a Novel Sequencing Device.* Nano Letters, 2010. **10**(9): p. 3237–3242.
14. Postma, H.W.C., *Rapid Sequencing of Individual DNA Molecules in Graphene Nanogaps.* Nano Letters, 2010. **10**(2): p. 420-425.
15. F. Traversi, et al., *Detecting the translocation of DNA through a nanopore using graphene nanoribbons.* Nat Nano, 2013. **8**(12): p. 939-945.
16. Spencer, C. and W. Meni, *Challenges in DNA motion control and sequence readout using nanopore devices.* Nanotechnology, 2015. **26**(7): p. 074004.
17. Branton, D., et al., *The potential and challenges of nanopore sequencing.* Nat Biotech, 2008. **26**(10): p. 1146-1153.
18. Zwolak, M. and M.D. Ventra, *Colloquium: Physical approaches to DNA sequencing and detection.* Reviews of Modern Physics, 2008. **80**(1): p. 141-165.
19. Georgakilas, V., et al., *Functionalization of Graphene: Covalent and Non-Covalent Approaches, Derivatives and Applications.* Chemical Reviews, 2012. **112**(11): p. 6156-6214.
20. Prasongkit, J., et al., *Theoretical Study of Electronic Transport through DNA Nucleotides in a Double-Functionalized Graphene Nanogap.* The Journal of Physical Chemistry C, 2013. **117**(29): p. 15421-15428.
21. Frisch, M.J., et al., *Gaussian 09.* 2009, Gaussian, Inc.: Wallingford, CT, USA.
22. Becke, A.D., *Density-functional thermochemistry. III. The role of exact exchange.* The Journal of Chemical Physics, 1993. **98**(7): p. 5648-5652.
23. Lee, C., W. Yang, and R.G. Parr, *Development of the Colle-Salvetti correlation-energy formula into a functional of the electron density.* Physical Review B, 1988. **37**(2): p. 785-789.

24. Breneman, C.M. and K.B. Wiberg, *Determining atom-centered monopoles from molecular electrostatic potentials. The need for high sampling density in formamide conformational analysis.* Journal of Computational Chemistry, 1990. **11**(3): p. 361-373.
25. Jorgensen, W.L. and J. Tirado-Rives, *The OPLS [optimized potentials for liquid simulations] potential functions for proteins, energy minimizations for crystals of cyclic peptides and crambin.* Journal of the American Chemical Society, 1988. **110**(6): p. 1657-1666.
26. Jorgensen, W.L., D.S. Maxwell, and J. Tirado-Rives, *Development and Testing of the OPLS All-Atom Force Field on Conformational Energetics and Properties of Organic Liquids.* J. Am. Chem. Soc., 1996. **118**(45): p. 11225–11236.
27. Lindsay, L. and D.A. Broido, *Optimized Tersoff and Brenner empirical potential parameters for lattice dynamics and phonon thermal transport in carbon nanotubes and graphene.* Physical Review B, 2010. **81**(20): p. 205441.
28. Smolyanitsky, A., *Molecular dynamics simulation of thermal ripples in graphene with bond-order-informed harmonic constraints.* Nanotechnology, 2014. **25**(48): p. 485701.
29. Horn, H.W., et al., *Development of an improved four-site water model for biomolecular simulations: TIP4P-Ew.* The Journal of Chemical Physics, 2004. **120**(20): p. 9665-9678.
30. Abascal, J.L.F., et al., *A potential model for the study of ices and amorphous water: TIP4P/Ice.* The Journal of Chemical Physics, 2005. **122**(23): p. 234511.
31. Van Der Spoel, D., et al., *GROMACS: Fast, flexible, and free.* Journal of Computational Chemistry, 2005. **26**(16): p. 1701-1718.
32. Hess, B., et al., *GROMACS 4: Algorithms for Highly Efficient, Load-Balanced, and Scalable Molecular Simulation.* Journal of Chemical Theory and Computation, 2008. **4**(3): p. 435-447.
33. Hatch, K., et al., *Demonstration that the shear force required to separate short double-stranded DNA does not increase significantly with sequence length for sequences longer than 25 base pairs.* Physical Review E, 2008. **78**(1): p. 011920.
34. Boland, T. and B.D. Ratner, *Direct measurement of hydrogen bonding in DNA nucleotide bases by atomic force microscopy.* Proceedings of the National Academy of Sciences of the United States of America, 1995. **92**(12): p. 5297-5301.
35. Minot, E.D., et al., *Tuning Carbon Nanotube Band Gaps with Strain.* Physical Review Letters, 2003. **90**(15): p. 156401.
36. Isacsson, A., *Nanomechanical displacement detection using coherent transport in graphene nanoribbon resonators.* Physical Review B, 2011. **84**(12): p. 125452.
37. Cosma, D.A., et al., *Strain-induced modifications of transport in gated graphene nanoribbons.* Physical Review B, 2014. **90**(24): p. 245409.
38. Datta, S., *Electron Transport in Mesoscopic Systems*. 1995, Cambridge: Cambridge University Press.
39. Pop, E., V. Varshney, and A.K. Roy, *Thermal properties of graphene: Fundamentals and applications.* MRS Bulletin, 2012. **37**(12): p. 1273-1281.
40. Smolyanitsky, A. and V.K. Tewary, *Manipulation of graphene's dynamic ripples by local harmonic out-of-plane excitation.* Nanotechnology, 2013. **24**(5): p. 055701.
41. Feng, Y., et al., *Nanopore-based Fourth-generation DNA Sequencing Technology.* Genomics, Proteomics & Bioinformatics, 2015. **13**(1): p. 4-16.
42. Schneider, G.F., et al., *Tailoring the hydrophobicity of graphene for its use as nanopores for DNA translocation.* Nat Commun, 2013. **4**.
43. Qi, J., et al., *Piezoelectric effect in chemical vapour deposition-grown atomic-monolayer triangular molybdenum disulfide piezotronics.* Nat Commun, 2015. **6**.

**Supplementary information**

**Nucleotide-functionalized graphene nanoribbons for accurate high-speed DNA sequencing**

Eugene Paulechka, Tsjerk A. Wassenaar, Kenneth Kroenlein, Andrei Kazakov, and

Alex Smolyanitsky

*S1. Strain-induced modification of electrical conduction in graphene nanoribbons*

The electronic properties of graphene nanoribbons (GNRs) in absence of externally induced strain depend on their width, aspect ratio, purity, edge type (zigzag or armchair) relative to the bias direction [1-6], and the presence of passivation at the edges [7], as well as the passivation content [7, 8]. Particularly relevant to our discussion, the GNR geometry determines the effective operating bias points suitable for strain detection when coherent transport dominates [1]. Further, geometrically perfect edges without chemical passivation are currently unlikely to be obtained experimentally, and thus the appropriate measurement strategy must be determined for a particular GNR.

Nevertheless, the *relative variation* of the electrical current around the baseline values due to deflection-induced strains can be estimated at the order-of-magnitude level for idealized cases. Furthermore, the nanomechanical deflections reported in the main text are for a zigzag-edged GNR and are generally valid for armchair GNRs of similar dimensions. Therefore, our discussion of the strain value estimates is not necessarily limited to the particular GNR type used in our MD simulations. Further, as seen in Eq. (S4) below and mentioned in the main text, a similar nanomechanical response can be obtained from GNRs of varying length and width (provided some requirements on the aspect ratio are met), thus allowing a degree of freedom in varying the dimensions, crucial for the design of the GNR properties in absence of strain. Here we briefly present the basic mechanisms underlying the effect of uniaxial strains on the electronic properties of GNRs, while the numerical estimates are provided in section S2.

The resistance of a GNR at a given appropriately selected bias point, excluding contact resistance for clarity, in the thermally activated regime is approximated as [9]:

$$R = \frac{R_0}{|t|^2}\left(1 + e^{\frac{E_{gap}}{kT}}\right), \qquad (S1)$$

where $R_0$ is the quantum resistance unit, $|t|^2$ is the effective transmission probability for electrons with a given energy $E$ (as dictated by the bias), such that $|E - E_F| > E_{gap}$ ($E_F$ and $E_{gap}$ are the Fermi level and the bandgap, as determined by the GNR dimensions, edge, *etc.*, respectively), and $T$ is the temperature. In the thermally activated conduction regime expected to dominate the water-immersed GNR at room temperature, the effect of strain is primarily due to modification of the number of carriers proportional to $e^{-\frac{E_{gap}}{kT}}$ via strain-induced change of $E_{gap}$, resulting in $\frac{\delta R}{R} \approx \frac{\delta E_{gap}}{kT}$ (independent of $E_{gap}$ itself in the perturbative approximation), where $\delta E_{gap} \approx 3t_0\varepsilon$, ($t_0 \approx 2.7\ eV$ is the nearest-neighbor electron hopping energy for graphene, $\varepsilon$ is the strain), thus yielding $\frac{\delta R}{R} \approx \frac{3t_0\varepsilon}{kT}$. The theoretical tight-binding estimate $\delta E_{gap} \approx 3t_0\varepsilon$ is close to the results obtained with density functional theory calculations for both zigzag and armchair GNRs under uniaxial strain [5].

For completeness, in the $T = 0$ K limit, Eq. (S1) is effectively replaced by the coherent transport term:

$$R = \frac{R_0}{|t|^2} \qquad (S2)$$

and the effect of strain is via modification of the effective (quantized) transmission probability $|t|^2$. In this case, the effect of strain on a gateless GNR is negligible [10]. However, around an appropriately selected bias point, strain can indeed be detected in an interferometer-type measurement setup with a relative variation of R estimated at $\frac{\delta R}{R} = -\frac{\delta|t|^2}{|t|^2} \approx \frac{h^2}{La_0}$, where $h$ is the out-of-plane deflection, $L$ is the effective GNR length, and $a_0$ is the C-C interatomic distance in graphene [10].

In the next sections we estimate the out-of-plane deflection $h$ and evaluate the order of magnitude of the changes in resistance induced by the strains due to forces expected in our system.

*S2. GNR deflection with lateral pre-strain*

As a rough estimate, the maximum out-of-plane deflection $h$ of an edge-clamped GNR of length $L$ due to force $F$ applied at $L/2$ is the solution of the following cubic equation:

$$F = \frac{2E_{2D}wh}{L^2}\left(\varepsilon_0 L + \frac{2h^2}{L}\right), \quad (S3)$$

where $E_{2D}$, $w$, and $\varepsilon_0$ are the two-dimensional Young's modulus of graphene, GNR width, and the initial pre-strain along the GNR length, respectively. A reasonable agreement with the simulated data was obtained with $E_{2D} \approx (E_{3D}h_0) = 106\ N/m$, where $E_{3D} = 1.06\ TPa$ and $h_0 = 0.1\ nm$ are the 3-D Young's modulus of graphene and its effective "continuum" thickness, respectively [11].

For $L$ = 15.5 nm, $F$ = 75 pN, and $\varepsilon_0$ = 0.5 %, Eq. (S3) yields $h$ = 2.24 Å, in reasonable agreement with the results in Figs. 3 (a, b) in the main text (results consistent with Fig. 3 (a, b) were also obtained in MD simulations with the DNA atomic charges set according to AMBER [12]). Without pre-strain ($\varepsilon_0 = 0$), the central deflection is:

$$h = L \left(\frac{F}{4E_{2D}w}\right)^{1/3}, \qquad (S4)$$

yielding $h_{\varepsilon_0=0} = 5.27$ Å. The deflection-induced strain in this case is $\varepsilon \approx 2\left(\frac{h}{L}\right)^2 = 0.23\ \%$, which causes an estimated $\frac{\delta R}{R} \approx \frac{\delta E_{gap}}{kT} = \frac{3t_0\varepsilon}{kT} = 71\%$ and $\frac{\delta R}{R} \approx \frac{h^2}{La_0} = 12\%$, according to Eqs. (S1) and (S2), respectively. For the A-T binding with a critical force of $F = 50$ pN, $h_{\varepsilon_0=0} = 4.60$ Å, yielding the $\frac{\delta R}{R}$ estimates of 32 % and 5.3 %, according to Eqs. (S1) and (S2), respectively.

With the experimentally obtained $E_{2D} = 352\ N/m$ [13], Eq. (S3) underestimates the deflections obtained in our simulations with $\varepsilon_0 = 0.5\ \%$. However, with lower $\varepsilon_0$, it yields deflections of comparable magnitude, and thus all of the estimates made here remain valid. For example, with $E_{2D} = 352\ N/m$ and $\varepsilon_0 = 0.1\ \%$, we obtain a deflection value of 1.9 Å for the C-G pair, and thus identical $\frac{\delta R}{R}$ estimates. Without pre-strain, the maximum deflection according to Eq. (S2) is $h_{\varepsilon_0=0} = 3.1$ Å, and thus $\frac{\delta R}{R} \approx \frac{\delta E_{gap}}{kT} = \frac{3t_0\varepsilon}{kT} = 24\%$ from Eq. (S1). Within the approximations made, all of these estimates are valid for GNRs of appropriately scaled dimensions.

*S3. Additional note on the GNR edge effects*

A very narrow GNR, such as the one used in our simulations, would present a region of locally reduced conductance in the nanopore region, given the closeness of the pore edge to the GNR edge. An additional effect on the electrical conductivity thus would arise from the local strain inhomogeneity near the pore, similar to the effects of local inhomogeneous strains shown elsewhere [1, 14]. Although only demonstrated in vacuum at zero temperature, the electronic properties of a narrow GNR could also be affected by the fact of nucleobase presence in the pore, even without functionalization [15]. However, these effects would be virtually non-existent in a considerably wider GNR at a finite temperature, given that the pore diameter would remain the same. Interestingly, a contribution from the pseudomagnetic field effect could arise in addition to the effects discussed earlier[16] in a wide GNR deflected by an effectively point force at the center.

*S4. Effects of rippling*

It has been shown previously that the local modulation of the graphene's nearest-neighbor electron hopping energetics by flexural ripples [17] can be described by the emergence of a gauge field [18-20]. An estimate of the time-averaged effect of the ripples can be obtained from considering them as carrier scatterers, which leads to an overall increase of the electrical resistivity [21], in addition to the temporal modulation of the current. In the long-wave approximation, this excess resistivity $\rho_r$ increases with the rippling strength (*e.g.*, in terms of the mean-square out-of-plane displacement $\langle h^2 \rangle$), while its size-scaling properties depend on the rippling Fourier scaling law $h_q^2$ [21]. Here, we discuss the qualitative effect of FGNR rippling on $\rho_r$ during DNA translocation by considering the wave-vector distributions $h_q^2$ and the $\langle h^2 \rangle$ averages, as obtained during the passage of G and non-G residues through the FGNR (see section S5 for the calculation details). Shown in Fig. S1 are the $h_q^2$ distributions for the FGNR at $T = 300$ K, along with the raw rippling data in the corresponding inset. As shown, the distributions are similar during the passage of G and non-G residues, although *the rippling strength during G passage is consistently lower*. The latter is expected, because even the relatively faint lateral strain can significantly suppress thermal flexural fluctuations.[22] Direct calculations of $\langle h^2 \rangle$ ($\propto \int h_q^2 \, d\Omega_q$, where $d\Omega_q$ is an area element in the 2-D reciprocal space) confirm this observation, yielding a decrease from 1.63 Å$^2$ to 1.35 Å$^2$ during the passage of non-G and G, respectively. An accurate quantitative estimate of $\Delta\rho_r/\rho_r$ due to FGNR deflection induced strain would crucially depend on the dimensions, as well as the fabrication methods of an experimentally relevant GNR. However, the relatively high sensitivity of the ripple scattering mechanism to $\langle h^2 \rangle$ (and thus to deflection-induced strain) can be revealed via previously estimated $\rho_r \propto n_r/n_c$ [21], where $n_r \propto \langle h^2 \rangle^2$ and $n_c \propto e^{-\frac{E_{gap}}{kT}}$ is the effective sheet density of the scatterers and charge carriers, respectively. As a result of excess strain due to G-induced FGNR deflection, $\frac{\Delta\rho_r}{\rho_r} \propto \frac{\Delta n_r}{n_r} - \frac{\Delta n_c}{n_c}$. Here, $n_r \propto \langle h^2 \rangle^2$ is considerably reduced (by ~30 %, from the $\langle h^2 \rangle$ estimates above) and $\Delta n_c/n_c \propto -\frac{3t_0\varepsilon}{kT} < 0$ due to strain-induced bandgap modulation, estimated at ~10 % above. The net result of this competition between strain-induced decreased scattering and a decrease in the number of charge carriers is reduction of $\rho_r$ by ~20 %. Therefore, if ripple scattering is expected to significantly contribute to the overall resistance in a given GNR, the described effect of strain-induced ripple suppression may become an additional mechanism contributing to the net current variation.

*S5. Out-of-plane rippling statistics*

For the *t*-th MD frame, an individual $h_q^2$ distribution was calculated directly from the atomic population of the FGNR as the corresponding 2-D Fourier transform of $(z_i - \bar{z}_t)^2$, where $z_i$ is the *i*-th atom's position along Z and $\bar{z}_t$ is the local plane level at time *t*. The distributions $h_q^2$ were presented as *averages of distributions* over multiple frames for each translocation portion (G and non-G), similarly to the statistical data presented elsewhere [17, 23, 24].

The *t*-th per-frame average from *N* atoms in the GNR is

$$\langle h_t^2 \rangle = \frac{1}{N-1} \Sigma_N (z_i - \bar{z}_t)^2, \qquad (S5)$$

and the grand average per multiple frames is calculated as $\langle h^2 \rangle = \frac{1}{\tau} \Sigma_\tau \langle h_t^2 \rangle$. Note that for a membrane deflected at the center, the use of a global "plane level" $\bar{z}_t = \frac{1}{N} \Sigma_N z_i$ is incorrect. Therefore, we used the local plane level $\bar{z}_{t,i}$ equal to the *per-atom* running time-average obtained from an infinite impulse response (IIR) filter. Ripple suppression was independently confirmed by using Eq. (S5), while calculating $\bar{z}_{t,i}$ from a second-order polynomial surface fit at every MD frame. The data in Fig. S1 and the grand averages discussed in section S4 were calculated over $\tau$ = 20 ns long periods of G and non-G translocation (see inset of Fig. S1; translocation data from Fig. 3 (b)). The frame spacing was 50 ps, resulting in a total of 400 frames used in the averaging for each passage.

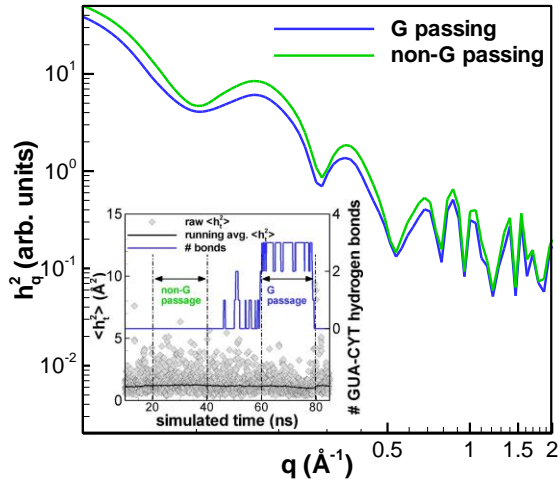

Figure S1. FGNR rippling distributions $h_q^2$ during the passage of G and non-G nucleobases. The inset shows the averaging regions, the raw $\langle h_t^2 \rangle$ data (as defined in section S4), as well as the $\langle h_t^2 \rangle$ running average data.

## References

1. Cosma, D.A., et al., *Strain-induced modifications of transport in gated graphene nanoribbons.* Physical Review B, 2014. **90**(24): p. 245409.
2. Nakada, K., et al., *Edge state in graphene ribbons: Nanometer size effect and edge shape dependence.* Physical Review B, 1996. **54**(24): p. 17954-17961.
3. Barone, V., O. Hod, and G.E. Scuseria, *Electronic Structure and Stability of Semiconducting Graphene Nanoribbons.* Nano Letters, 2006. **6**(12): p. 2748-2754.
4. Yoon, Y. and J. Guo, *Effect of edge roughness in graphene nanoribbon transistors.* Applied Physics Letters, 2007. **91**(7): p. 073103.
5. Li, Y., et al., *Strain effects in graphene and graphene nanoribbons: The underlying mechanism.* Nano Research, 2010. **3**(8): p. 545-556.
6. Son, Y.-W., M.L. Cohen, and S.G. Louie, *Energy Gaps in Graphene Nanoribbons.* Physical Review Letters, 2006. **97**(21): p. 216803.
7. Lu, Y.H., et al., *Effects of edge passivation by hydrogen on electronic structure of armchair graphene nanoribbon and band gap engineering.* Applied Physics Letters, 2009. **94**(12): p. 122111.
8. Zeng Yong-Chang, T.W., Zhang Zhen-Hua, *Electronic properties of graphene nanoribbons with periodical nanoholes passivated by oxygen.* Acta Physica Sinica, 2013. **62**(23): p. 236102-236102.
9. Minot, E.D., et al., *Tuning Carbon Nanotube Band Gaps with Strain.* Physical Review Letters, 2003. **90**(15): p. 156401.
10. Isacsson, A., *Nanomechanical displacement detection using coherent transport in graphene nanoribbon resonators.* Physical Review B, 2011. **84**(12): p. 125452.

11. Scarpa, F., S. Adhikari, and A.S. Phani, *Effective elastic mechanical properties of single layer graphene sheets.* Nanotechnology, 2009. **20**(6): p. 065709.
12. Duan, Y., et al., *A point-charge force field for molecular mechanics simulations of proteins based on condensed-phase quantum mechanical calculations.* Journal of Computational Chemistry, 2003. **24**(16): p. 1999-2012.
13. Booth, T.J., et al., *Macroscopic Graphene Membranes and Their Extraordinary Stiffness.* Nano Letters, 2008. **8**(8): p. 2442-2446.
14. Smolyanitsky, A. and V.K. Tewary, *Simulation of lattice strain due to a CNT–metal interface.* Nanotechnology, 2011. **22**(8): p. 085703.
15. Nelson, T., B. Zhang, and O.V. Prezhdo, *Detection of Nucleic Acids with Graphene Nanopores: Ab Initio Characterization of a Novel Sequencing Device.* Nano Letters, 2010. **10**(9): p. 3237–3242.
16. Fogler, M.M., F. Guinea, and M.I. Katsnelson, *Pseudomagnetic Fields and Ballistic Transport in a Suspended Graphene Sheet.* Physical Review Letters, 2008. **101**(22).
17. Fasolino, A., J.H. Los, and M.I. Katsnelson, *Intrinsic ripples in graphene.* Nature Materials, 2007. **6**(11): p. 858 - 861.
18. Morozov, S.V., et al., *Strong Suppression of Weak Localization in Graphene.* Physical Review Letters, 2006. **97**(1): p. 016801.
19. Eun-Ah, K. and A.H.C. Neto, *Graphene as an electronic membrane.* EPL (Europhysics Letters), 2008. **84**(5): p. 57007.
20. Abedpour, N., et al., *Roughness of undoped graphene and its short-range induced gauge field.* Physical Review B, 2007. **76**(19): p. 195407.
21. Katsnelson, M.I. and A.K. Geim, *Electron scattering on microscopic corrugations in graphene.* Philosophical Magazine, 2008. **366**(1863): p. 195-204.
22. Gao, W. and R. Huang, *Thermomechanics of monolayer graphene: Rippling, thermal expansion and elasticity.* Journal of the Mechanics and Physics of Solids, 2014. **66**: p. 42-58.
23. Los, J.H., et al., *Scaling properties of flexible membranes from atomistic simulations: Application to graphene.* Physical Review B, 2009. **80**(12): p. 121405.
24. Smolyanitsky, A., *Effects of thermal rippling on the frictional properties of free-standing graphene.* RSC Advances, 2015. **5**(37): p. 29179-29184.